\documentclass[preprint,showpacs,preprintnumbers,amsmath,amssymb]{revtex4}

\usepackage{graphicx}% Include figure files
\usepackage{dcolumn}% Align table columns on decimal point
\usepackage{bm}% bold math

\begin{document}

\title{Thermal Spin-Accumulation in Electric Conductors and Insulators}
 \author{J.-E. Wegrowe} \email{jean-eric.wegrowe@polytechnique.fr} \author{H.-J. Drouhin} 
\affiliation{Ecole Polytechnique, LSI, CNRS and CEA/DSM/IRAMIS, Palaiseau F-91128, France}
\author{D. Lacour} 
\affiliation{Institut Jean Lamour, UMR CNRS 7198, Universit\'e H. Poincarr\'e, F -5406 Vandoeuvre les Nancy, France}

\date{\today}

\begin{abstract}
The interpretation of some recent measurements of spin-dependent voltage for which the electric conduction does not play a role rises some new fundamental questions about the effects of spin-dependent heat currents. A two spin-channel model is proposed in order to describe the effect of out-of-equilibrium spin-dependent heat carriers in electric conductors and insulators. It is shown that thermal spin-accumulation can be generated by the heat currents only over an arbitrarily long distance for both electric conductors or electric insulators. The diffusion equations for thermal spin-accumulation are derived in both cases, and the principle of its detection based on Spin-Nernst effect is described.
\end{abstract}

\pacs{72.25.Mk, 85.75.-d \hfill}

\maketitle

%\section{Introduction}

%\section{Introduction}

The role played by thermoelectric effects in the context of giant magnetoresistance (GMR) has long been recognized \cite{MTEP1,MTEP2,Shi,Piraux,Gravier,Fukushima} and exploited for possible applications in spintronics. Indeed, spin-dependent Seebeck and spin-dependent Peltier measurements are direct generalizations of giant magnetoresistance experiments performed in the presence of temperature gradient, that can be described in the framework of the two spin-channel model for conduction electrons \cite{Johnson, Wyder,Valet, PRB2000}. Starting from the coupled transport equations of thermoelectricity, the introduction of the spin-dependent conduction coefficients $\sigma_{\updownarrow}$ and the spin-dependent Seebeck cross-coefficients $\mathcal S_{\updownarrow}$ suffices to describe most of the spin-dependent thermoelectric effects in electric conductors \cite{Sinova,Tulapurkar,Bauer,Fabian,Uchida_theo,Nunner,MTEP,Gravier2}. However, recent experiments pioneered by the group of  Uchida et al. \cite{Uchida,Jaworsky,Uchida2,Sharoni,RezendePRL,Rezende} and still not well understood \cite{Sinova,Bauer,Fabian} opened the way to a new class of spintronics effects in which no electric current is flowing and only spin-dependent heat currents are present in the sample. Such experiments are performed in electric conductors in the so-called non-local geometry (Fig. 1) at a long distance from the current injection, or in electrical insulators (Fig. 2).

The goal of this work is to propose a phenomenological description of spin-accumulation produced by spin-dependent heat currents. The model first describes the spin-thermocouple effect that allows spin-accumulation to be generated by a temperature gradient in conductors over arbitrarily large distances. This spin-dependent version of the thermocouple effect (that allows a temperature difference to be measured with a voltmeter in an open circuit), is corollary of the Seebeck effect (that allows an electric current to be generated with a temperature difference in a closed circuit). Both effects are deduced from the same thermoelectric transport equations with different boundary conditions and different power dissipation.  The model is then applied to the case of electric insulators within the two spin-channel model of heat conduction. Finally, the transverse spin-dependent thermocouple effect occurring in the spin- Hall or spin-Nernst detector is also described. In all cases, the corresponding diffusion equations for the spin-accumulation are derived.

% \section{Spin-dependent thermoelectric Effects in non-local device at partial equilibrium}

\begin{figure}
\includegraphics[scale=0.5]{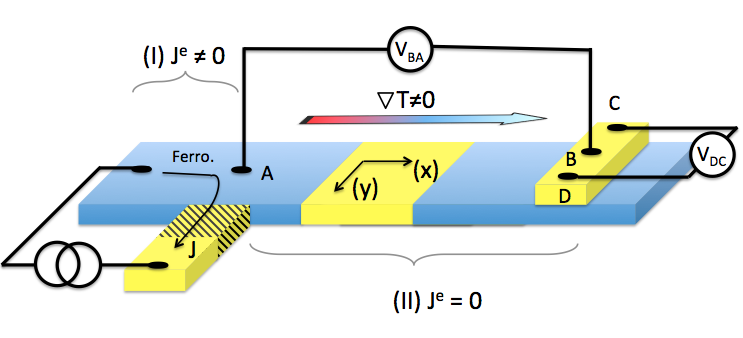}
\caption{Schematic view of the spintronics non-local device with temperature gradient. Branch (II) under interest constitutes an open circuit from the point of view of the electric charges ($J^e=0$) but not from the point of view of the heat currents. The voltage is measured between points $A$ and $B$ in branch (II), which contains ferromagnetic/non-ferromagnetic interfaces, or between the points $C$ and $D$ of the lateral electrode.}
\label{fig1}
\end{figure}

\begin{figure}
\includegraphics[scale=0.6]{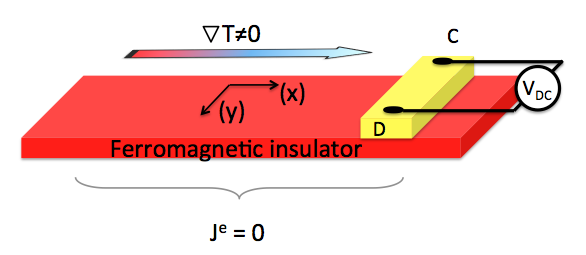}
\caption{Schematic view of non-local device with ferromagnetic insulator. Heat current is injected through the device and spin-accumulation can be produced in the lateral electrode. The potential due to spin-dependent Nernst effect is measured transversally along the $y$ direction.}
\label{fig2}
\end{figure}

For a  non-isothermal ferromagnetic material - that can be an electric conductor or an electric insulator - the system under interest is composed of a statistical ensemble of out-of-equilibrium heat carriers of two different species: the heat carriers that carry a up spin $\uparrow$ and the heat carriers that carry a down spin $\downarrow$. In the case of electric conductors, the spin-dependent heat carriers are also electric carriers, while for electric insulators, the spin-dependent heat carriers are spin-dependent excitation modes (e.g. magnons). In both cases however, the two sub-systems  of $\uparrow$ and  $\downarrow$ heat carriers are described by the corresponding local chemical potential $\mu_{\uparrow}$ and $\mu_{\downarrow}$, assuming the hypothesis of local equilibrium \cite{DeGroot}. In line with the description of the giant magnetoresistance \cite{Johnson,Wyder,Valet}, we define the {\it thermal spin-accumulation} by the difference of the two chemical potentials $\Delta \mu = \mu_{\uparrow} - \mu_{\downarrow}$ of the heat carriers.
In the framework of the non-equilibrium thermodynamic theory, the bivaluated spin variable $\updownarrow$ defines an internal degree of freedom \cite{DeGroot,PRB2000,Entropy}. The spin-relaxation occurring during the transport process is then treated as a chemical reaction that transforms heat carriers from the up channel to the down channel or inversely. The rate $\dot \psi^{th}$ of this reaction is a flux of spin variable in the spin space. Since this reaction changes the density of the heat carriers of the spin channels, the corresponding chemical affinity is $\Delta \mu$. The Onsager relation that links the flux $\dot \psi^{th}$ to the force $\Delta \mu$ takes the same form as for electric spin-accumulation \cite{PRB2000,Entropy,MTEP}: 
\begin{equation}
\dot \psi^{th} = L_{th} \Delta \mu,
\label{ChemReac}
\end{equation}
where we have introduced the phenomenological Onsager coefficient $L_{th}$ that describes the spin-dependent relaxation. At stationary regime (i.e. for the state of minimum entropy production of the system), the internal energy densities $u_{\uparrow}$ and $u_{\downarrow}$ obey the following conservation equations: 
\begin{equation}
\begin{array}{ccc}
\frac{\partial u_{\uparrow}}{\partial t}  = -  \vec{\nabla}  . \vec J_{\uparrow}^q  - \dot \psi^{th} + \mathcal R/2 & = & 0\\
\frac{\partial u_{\downarrow}}{\partial t}  =  -   \vec{\nabla} . \vec J^q_{\downarrow} + \dot \psi^{th} + \mathcal R/2 & = & 0,
\end{array}
\label{ThermalCons}
\end{equation}
 where $\mathcal R$ describes the dissipation out of the system (e.g. by radiation) and $\dot \psi^{th}$ describes the spin relaxation from one heat conduction channel to the other.\\

In the case of {\it electric conductors} and in a first approximation, the heat carriers and the electric carriers are the same, so that the chemical potentials of the heat conduction channels is also that of the electric conduction channels. The well-known thermoelectric transport equations relate the electric currents $\vec J^e_{\updownarrow}$ and the heat currents $\vec J^q_{\updownarrow}$ to the corresponding forces $\vec \nabla \tilde \mu_{\updownarrow}$ and $\vec \nabla T$  \cite{MTEP,Nunner,McGraw}:

\begin{equation}
%\begin{array}{ccc}
\vec J^{e}_{\updownarrow} = - \frac{\sigma_{\updownarrow}}{e} \, \vec \nabla \tilde \mu_{\updownarrow} 
- \frac{\sigma_{\updownarrow} \mathcal S_{\updownarrow}}{e} \,  \vec \nabla T,
\label{ThermoCurrentA}
\end{equation}
\begin{equation}
\vec J^{q}_{\updownarrow} =     \lambda_{\updownarrow} \, \vec \nabla T - \frac{\Pi_{\updownarrow}}{e} \, \vec \nabla \tilde \mu_{\updownarrow}
%\end{array}
\label{ThermoCurrentB}
\end{equation}
where $ \lambda_{\updownarrow}$ is the spin-dependent Fourier coefficient and $ \Pi_{\updownarrow} = \sigma_{\updownarrow} \mathcal S_{\updownarrow}T$ is the spin-dependent Peltier coefficient.  The electrochemical 
potential reads $\tilde \mu_{\updownarrow} = 
\mu_{ch,\updownarrow} + eV$ where $e$ is the charge of the electric carriers and $V$ the electric voltage.  We assume that the temperature dependence of the transport 
coefficients is negligible. 

The condition for the partial equilibrium $\vec J_{\updownarrow} = 0$ is then 
given by 
\begin{equation}
\vec \nabla \tilde \mu_{\updownarrow} = - \mathcal S_{\updownarrow} \vec \nabla  T
\label{PartEquil2}
\end{equation}
 Eqs. (\ref{PartEquil2}) can be reformulated with introducing the total Seebeck coefficient
$\mathcal S_0 =  \mathcal S_{\uparrow} + 
\mathcal S_{\downarrow}$:
\begin{equation}
 \vec \nabla \left (\tilde \mu_{\uparrow} +  \tilde \mu_{\downarrow} \right )
= -  \mathcal S_{0} \, \vec \nabla T
\label{PartEqA}
\end{equation}
and the Seebeck asymmetry coefficient $ \Delta \mathcal S = \mathcal S_{\uparrow} - \mathcal S_{\downarrow}$: 
\begin{equation}
\vec \nabla (\Delta \mu) = 
- \Delta \mathcal S  \, \vec \nabla T.
\label{PartEqB}
\end{equation}
Eq. (\ref{PartEqA}) is nothing but the equation of the thermocouple. Equation (\ref{PartEqB}) is, in contrast, a direct consequence of the presence of spin-dependent Seebeck coefficients, and describes the {\it spin-dependent thermocouple effect}. Accordingly, it is expected that {\it a gradient of temperature produces a spin-accumulation gradient without electric current and without non-local spin injection}. The usual thermocouple effect allows the spin-accumulation generated by the temperature gradient to be measured:  inserting Eq. (\ref{PartEqB}) in to Eq. (\ref{PartEqA}) we obtain:
\begin{equation}
\vec \nabla \left (\tilde \mu_{\uparrow} +  \tilde \mu_{\downarrow} \right )
 = 
\frac{\mathcal S_{0}}{\Delta \mathcal S}  \, \vec \nabla (\Delta \mu).
\label{Result}
\end{equation}
Eq. (\ref{Result}) is able to account for the observed Spin- Seebeck effects measured without electric current ($\vec J^e_{\uparrow} = \vec J^e_{\downarrow} = 0$) at a distance arbitrarily large in the configuration of Fig. 1. The integration between the points $A$ and $B$ leads to the spin-dependent potential measured:
\begin{equation}
V_B -  V_A 
 = 
\int_A^{B} \frac{\mathcal S_{0}}{\Delta \mathcal S}  \, \frac{\partial \Delta \mu}{\partial x} dx.
\label{Result_Int}
\end{equation}

On the other hand, Eqs. (\ref{ThermoCurrentA}) to Eq. (\ref{Result}) can be generalized with the introduction of tensorial transport coefficients $\{ \sigma_{\updownarrow} \}_{ij}$, $\{ \mathcal S_{\updownarrow}\}_{ij}$ ($i,j = \{x,y,z\}$ ) where the non-diagonal elements $i \ne j$ account respectively for the Hall effect  \cite{SpinHall} and the corresponding  Nernst effect in homogeneous isotropic materials \cite{DeGroot}. The detection of the transverse potential difference $V_{DC}$ performed on the right electrode in Fig. 1 and Fig. 2 is described by the transverse spin-dependent thermocouple effect, or spin-dependent Nernst effect. If $\Delta S_{xy} \ne 0$, the projection of Eq. (\ref{PartEqB})  over the $Oy$ axis gives, after integration:
 
\begin{equation}
V_D - V_C = \int_C^D \frac{\partial \Delta \mu}{\partial y} dy =  \int_C^D \Delta \mathcal S_{xy} \frac{\partial T}{\partial x} dy.
\label{Nernst}
\end{equation}

Eq. (\ref{Result_Int}) and Eq. (\ref{Nernst}) are analogous to that of the usual spin-accumulation calculation for the giant magnetoresistance effect, with the difference that the spin-accumulation is produced by the heat current instead of the electric current, and that the spin-accumulation $\Delta \mu(x)$ and $\Delta \mu(y)$ are no longer governed by the usual diffusion equation of the spin-accumulation generated by electric current injection \cite{Johnson,Wyder,Valet,PRB2000}.

 Defining the total heat flux along the $x$ axis (see Fig. 1 and Fig. 2) $J_0^q = J^q_{\uparrow} + J^q_{\downarrow}$ and he spin-dependent heat flux  $\Delta J^q = J^q_{\uparrow} - J^q_{\downarrow}$, we have from Eq. (\ref{ThermoCurrentB}):
\begin{equation}
\begin{array}{ccc}
J^{q}_0 & =&  \lambda_{0} \, \frac{\partial T}{\partial x} -  \frac{\Pi_0}{2e} \, \left ( \frac{ \partial \tilde \mu_{0}}{\partial x} + \beta \frac{ \partial \Delta \mu}{\partial x} \right)\\
\Delta J^{q} & = &  \Delta \lambda \, \frac{\partial T}{\partial x} -  \frac{\Pi_0}{2e} \, \left ( \frac{ \partial \Delta \mu}{\partial x} + \beta \frac{ \partial \tilde \mu_0}{\partial x} \right),
\end{array}
\label{ThermoCurrBIS}
\end{equation}
where $\tilde \mu_0 = \tilde \mu_{\uparrow} + \tilde \mu_{\downarrow}$, $\lambda_0 = \lambda_{\uparrow} + \lambda_{\downarrow}$,  $\Delta \lambda = \lambda_{\uparrow} - \lambda_{\downarrow}$, $\Pi_{0} = \Pi_{\uparrow}  + \Pi_{\downarrow}$,  and $\beta = \left (\Pi_{\uparrow} - \Pi_{\downarrow} \right )/ \left (\Pi_{\uparrow} + \Pi_{\downarrow} \right )$.

Note that for $J^{q}_0 = 0 $, the flow $\Delta J^q$ is a ``{\it pure spin-current}" \cite{Sinova,Bauer} that cannot be distinguished from its electric homologue $\Delta J^e = J^e_{\uparrow} - J^e_{\downarrow}$ with $J^e_0 = J^e_{\uparrow} + J^e_{\downarrow} = 0$ (and more generally, a ``pure spin-current"  is independent of the nature of the spin carriers involved).

Inserting Eq. (\ref{ThermoCurrBIS}) and Eq. (\ref{ChemReac}) into Eq. (\ref{ThermalCons}), we deduce the diffusion equation for $\Delta \mu$: 
\begin{equation}
\frac{\partial ^2 \Delta \mu}{\partial x^2} - \frac{\Delta \mu}{l_{c}^2} = C_1 \frac{\partial^2 T}{\partial x^2} + C_2,
\label{difusionEq}
\end{equation}
where we have introduced the thermal spin-diffusion length for conducting materials 
\begin{equation}
l_{c} = \sqrt{\frac{\Pi_0 \left (1 - \beta^2 \right)}{4 e L_{th}}}
\end{equation}
and the constants $ C_1 = 2 e \left (\Delta \lambda - \lambda_0 \beta \right )/\left ( \Pi_0 (1 - \beta^2)\right ) $, $ C_2 = 2\beta \mathcal R e/\left ( \Pi_0 (1 - \beta^2) \right)$.

Assuming a uniform temperature gradient and negligible external dissipation $\mathcal R \approx 0$, the solutions of Eq. (\ref{difusionEq}) have the form $\Delta \mu(x)= \Delta \mu(0) \left ( a e^{x/l_{th} } + b e^{-x/l_{th}} \right )$ where $a$ and $b$ are constants imposed by the boundary conditions. It is the same solutions as that of the usual spin-accumulation responsible for GMR effects, with the difference that the diffusion length is defined with the help of the Peltier asymmetry coefficient $\beta = \Delta \Pi/ \Pi_0$ instead of the asymmetry of the electric conductivity,  and $L_{th}$ plays the role of the usual spin-flip Onsager coefficient  \cite{PRB2000,Tulapurkar}.\\

In the case of spintronic devices with {\it electric insulators} (Fig. 2), there are no electric carriers and the chemical potentials $\mu^{th}_{\updownarrow}$ are that of the corresponding heat carriers only (magnon, spin-dependent phonon, etc). The heat current is composed of a drift part $\vec J^q_{drift \, \updownarrow}= \lambda_{\updownarrow} \vec \nabla T$ and a diffusion part $\vec J^{q}_{diff \,\updownarrow} = -\mathcal L_{\updownarrow} \vec \nabla \mu^{th}_{\updownarrow}$, where $\mathcal L_{\updownarrow}$ is the corresponding Onsager coefficient related to the diffusion coefficient $D_{\updownarrow} = kT \mathcal L_{\updownarrow}/n^{th}$, where $n^{th}$ is the density of the heat carriers (for the sake of simplicity, we assume isotropic diffusion). In the configuration of Fig. 2, the transport equations reads:
\begin{equation}
\begin{array}{cc}
J^{q}_{\uparrow} = \lambda_{\uparrow} \, \frac{\partial T}{\partial x} -  \mathcal L_{\uparrow}  \, \frac{ \partial \mu^{th}_{\uparrow}}{\partial x}\\
J^{q}_{\downarrow} = \lambda_{\downarrow} \, \frac{\partial T}{\partial x} -  \mathcal L_{\downarrow}  \, \frac{ \partial \mu^{th}_{\downarrow}}{\partial x}.
\end{array}
\label{ThermoCurrent2}
\end{equation}
The diffusion equation for the chemical potential difference (i.e. spin accumulation) is deduced by inserting  Eq. (\ref{ThermoCurrent2}) and Eq. (\ref{ChemReac}) into Eqs. (\ref{ThermalCons}):
 \begin{equation}
\frac{\partial^2 \Delta \mu^{th}}{\partial x^2}  - \frac{\Delta \mu^{th}}{l_{in}^2} = \tilde C_1 \frac{\partial^2 T}{\partial x^2} + \tilde C_2,
\label{Diff3}
\end{equation}
where we have introduced the thermal diffusion length for electric insulators 
\begin{equation}
l_{in} =  \sqrt{ \frac{\mathcal L_0 \left ( 1- \tilde \beta^2 \right) }{4 L_{th}} },
\end{equation}
where $\mathcal L_0 = \mathcal L_{\uparrow} +  \mathcal L_{\downarrow}$ and $ \tilde \beta =  \left ( \mathcal L_{\uparrow} - \mathcal L_{\downarrow} \right )/\mathcal L_0$. On the other hand, 
$ C_1 = 2 (\Delta \lambda - \lambda_0 \tilde \beta ) / \left( \mathcal L_0 (1 - \tilde \beta^2) \right) $ and $\tilde C_2 = 2 \tilde \beta \mathcal R /((1 - \tilde \beta^2) \mathcal L_0)$.

This spin-accumulation in the electrical insulator is transferred to a conducting electrodes through the condition of continuity of the heat currents at the interface. In that case also, the final profile of the spin-accumulation throughout the insulator/conductor interface is similar to that of the usual GMR spin-accumulation under electric current.\\ 

In conclusion, the two spin-channel model has been applied to spin-dependent heat transport, in a circuit that is closed from the point of view of heat currents, and open for the point of view of electric currents. The result of the model shows that the heat current is able to play the same role as the electric current in a usual ``local" spin-valve system, whatever the spin-dependent heat carriers considered. In particular, spin-injection and spin-accumulation can be performed with spin-dependent heat currents only at an interface. The model explains the recent observations of spin-dependent voltage measured in electric conductors of electric insulators, in the configurations depicted in Fig. 1 and Fig. 2.

 \end{document}